\documentclass[%
 aip,
 amsmath,amssymb,
 reprint,%
]{revtex4-2}

\usepackage{graphicx}
\usepackage{dcolumn}
\usepackage{bm}

\usepackage[utf8]{inputenc}
\usepackage[T1]{fontenc}
\usepackage{mathptmx}
\usepackage{etoolbox}

\usepackage{color}
\usepackage{verbatim}
\usepackage[caption=false]{subfig}

\makeatletter
\def\@email#1#2{%
 \endgroup
 \patchcmd{\titleblock@produce}
  {\frontmatter@RRAPformat}
  {\frontmatter@RRAPformat{\produce@RRAP{*#1\href{mailto:#2}{#2}}}\frontmatter@RRAPformat}
  {}{}
}%
\makeatother
\begin{document}

\preprint{AIP/123-QED}

\title[Projected Density Matrix]{The Polarization Projected Density Matrix: A Practical Way to Recover Molecular Frame Information from Isotropic Samples}
\author{R. L. Thurston}
\affiliation{Lawrence Berkeley National Lab - Chemical Sciences Division}

\author{Niranjan Shivaram}
\affiliation{Department of Physics and Astronomy, Purdue University, West Lafayette, IN 47907, United States}
\affiliation{Purdue Quantum Science and Engineering Institute, Purdue University, West Lafayette, IN 47907, United States}

\author{Th. Weber}
\affiliation{Lawrence Berkeley National Lab - Chemical Sciences Division}

\author{L. Z. Tan}
\affiliation{Lawrence Berkeley National Lab - Molecular Foundry}

\author{D. S. Slaughter}
\affiliation{Lawrence Berkeley National Lab - Chemical Sciences Division}
\email{DSSlaughter@lbl.gov}

\date{\today}

\begin{abstract}
We present a novel approach to model ultrafast time-dependent nonlinear optical polarization sensitive signals emitted from randomly-oriented molecules. By projecting the laboratory-frame analyzer polarization axis into the molecular-frame and linking that axis with the density matrix through a tensor product, we demonstrate an approach to find a specific molecular orientation that yields a good approximation to simulated four-wave mixing signals produced by the same model but with averaging over molecular orientation. 
\end{abstract}

\maketitle

\section{\label{sec:level1}Introduction}

Nonlinear spectroscopic methods have provided rich and deep insight into electronic and nuclear dynamics in atomic and molecular systems. Recent examples include the observation of long range ultrafast energy transfer between optically active donor and acceptor molecules, which were excited in a microcavity \cite{russo_direct_nodate}, a study of charge transfer dynamics between electron donor and acceptor regions in oligomers \cite{gobeze_characterization_2024}, and the observation of phonon softening in resonant nanoelectromechanical systems \cite{yang_raman_2022}. The analysis of the measured spectra require sophisticated theoretical treatments. Many approaches to simulate nonlinear spectra using any of several quantum chemistry packages\cite{aidas_dalton_2014, shao_advances_2015} offer the ability to compute frequency domain nonlinear responses to the desired order. Such procedures are usually computationally intensive depending on the the size and complexity of the molecular system.

Recent advances in the computation of nonlinear time-domain resolved spectra show promising advantages in the interpretation of ultrafast molecular dynamics. Rose, and Krich \cite{rose_numerical_2019} have implemented an efficient means of simulating time domain nonlinear spectra with the Ultrafast Ultrafast (UF\textsuperscript{2}) software package, which simulates nonlinear spectra using a Fourier-based approach with a Runge–Kutta–Euler direct propagation method to estimate the spectroscopic signals from the Liouville equation. They also demonstrated the ability to automatically compute signals from all Feynman diagrams up to the 5\textsuperscript{th} order \cite{rose_automatic_2021} and treat the case of an open quantum system \cite{rose_efficient_2021}. Other simulation methods, such as those based on the nonlinear-response-function (NRF) approach \cite{gelin_equation--motion_2022}, have been developed and applied toward predicting nonlinear signals that are due to nonadiabatic dynamics in molecular systems in the XUV and X-Ray regimes \cite{kowalewski_simulating_2017}. Taken together, the collection of methods and packages are both powerful and general. However, these approaches do not typically include polarization sensitivity in the quantum level simulation and can become significantly more expensive under orientational averaging.

Polarization spectroscopies measure the signal fields with one or more specified polarization direction, which in turn specifies the physically relevant information to be measured. In particular, polarization sensitive spectroscopies are established to be sensitive to molecular orientations \cite{gelin_equation--motion_2022, gaynor_multimode_2021, tokmakoff_orientational_1996, woutersen_structure_2000}. By explicitly connecting polarization sensitivity into a nonlinear time-domain simulation method, we aim to access molecular frame information, such as electronic motion in the molecular frame, using these polarization sensitive techniques. 

One consequence of the sensitivity of nonlinear polarization to molecular orientation that relates to isotropic samples, as discussed by Gelin et al. \cite{gelin_equation--motion_2022}, is that orientational averaging must be done in order to properly represent an observed polarization sensitive signal. While not unique to polarization spectroscopy \cite{kwak_rigorous_2015, andrews_using_2004}, as many spectroscopic signals are orientation dependent, this requirement of orientational averaging, when considering spectroscopic signals of isotropic samples, does impose extra computational cost on any method that is trying to simulate spectra of isotropic samples. For the case of a third order polarization calculation, which is typically used to describe time resolved optical Kerr effect spectroscopy \cite{boyd_non-linear_2008, palese_femtosecond_nodate, mcmorrow_femtosecond_1988}, a general method to account for orientational averaging requires sampling a minimum of 27 different orientations \cite{gelin_equation--motion_2022}. If all of the 81 different tensor elements of the 3\textsuperscript{rd}-order susceptibility are calculated, 81 different orientations may be required\cite{andrews_using_2004, boyd_non-linear_2008}. Consequently, orientational averaging adds a minimum factor of 27 to the cost of 3\textsuperscript{rd}-order spectroscopy simulations.

We present a method to recover molecular frame information for a sample of randomly-oriented molecules to identify a single representative orientation of a molecular target that generates approximately the same signal as is achieved through orientation averaging. To find the representative orientation, we introduce the projected density matrix, an object containing information about both the molecular frame density matrix and the laboratory frame measurement axis. A consequence of this implicit representation of the molecular and laboratory frames is that the projected density matrix allows for perturbative simulations to retain more detailed quantum information that may otherwise be lost by orientation averaging. 

\begin{figure*}
    {\includegraphics{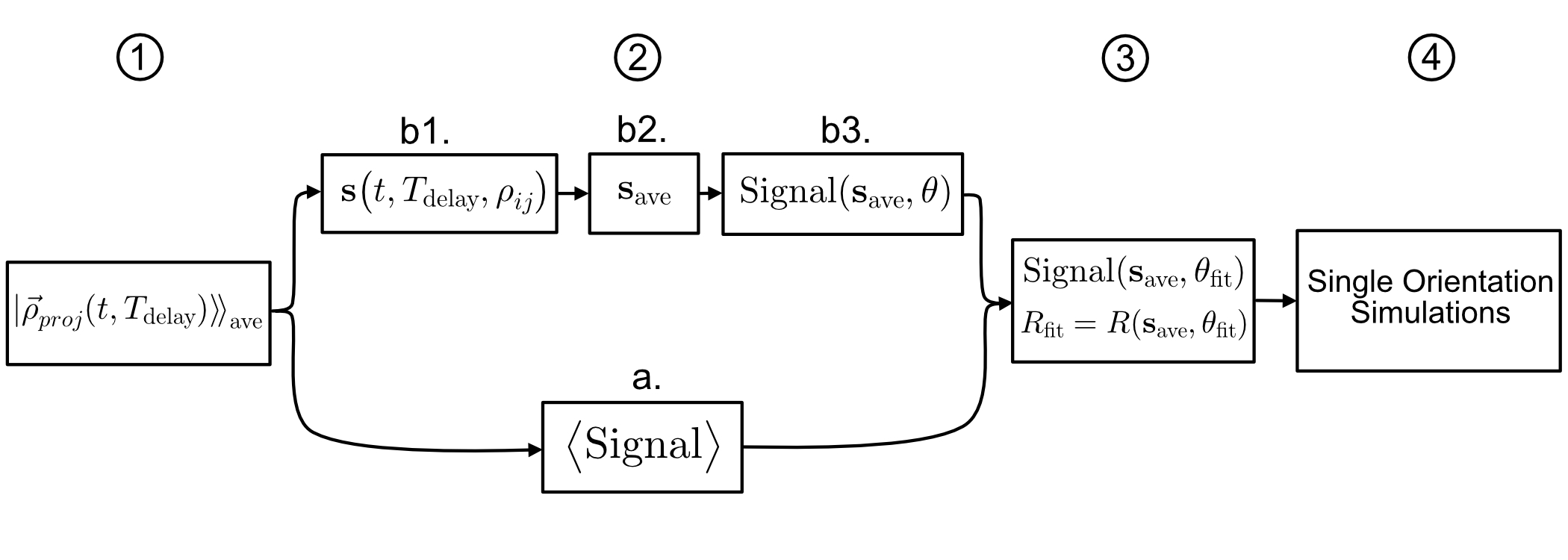}}
    \caption{Workflow diagram illustrating the steps used when simulating orientationally averaged nonlinear spectroscopic signals using a representative orientation given an orientationally averaged calculation. (1) First an orientationally averaged calculation over a restricted simulation time window given the desired order of perturbation theory is performed.  (2b1) This is used to compute the molecular frame polarization analyzer axis $\mathbf{s}$ for each moment in simulation time ($t$), for each density matrix element ($\rho_{ij}$), and for each time-delay ($T_{\text{delay}}$) between the incident pulses. (2b2) For each time-delay, we compute an average $\mathbf{s}_{\text{ave}}$ that approximates the time-integrated system response by averaging $\mathbf{s}$ over simulation time and over each density matrix element. (2b3) Using the averaged $\mathbf{s}_{\text{ave}}$, a single orientation fit of the signal is performed in which the computed time-delay-dependent signals from the orientationally averaged calculation (2a) and the single orientation calculation are compared. To perform the fit, we use the $\mathbf{s}_{\text{ave}}$ and samples of the remaining independent angle to define a set of molecular orientations that are used in the comparison. (3) Using the best sampled fit, we can lastly use the recovered orientation, which we use as an approximation for orientational averaging, to extend the simulation window (4).
    }\label{sec_sim:fig:workflow_for_single_run}
\end{figure*}

We apply the following procedure illustrated in Fig.~\ref{sec_sim:fig:workflow_for_single_run} to approximate orientationally averaged nonlinear polarization calculations using a single molecular orientation. First, we perform an orientationally averaged calculation over a narrow simulation time window using the projected density matrix. We then compute signals for a series of fixed orientations, comparing each signal with the orientation-averaged result. A representative orientation is found if the corresponding signal agrees with the orientation-average. Under the assumption that this representative orientation is a valid approximation of the averaged signal over extended time-windows, we expand the real-time simulation and thus expand the time-delay window of the simulation with a significant computational cost reduction offered by the single geometry. The approach enables time-domain calculations that are directly comparable with experiments, while still accounting for orientational averaging. After presenting the theoretical framework, we implement the method, apply it to the simulation of an electronic model of nitrobenzene, and demonstrate the capabilities with time resolved optical Kerr effect simulations to investigate the dependence of the retrieved representative molecular orientation on the applied pulse durations.

\begin{figure}[]
    \centering
     {\includegraphics[width=0.9\columnwidth,trim={1.0cm 1.0cm 1.0cm 0.5cm},clip]{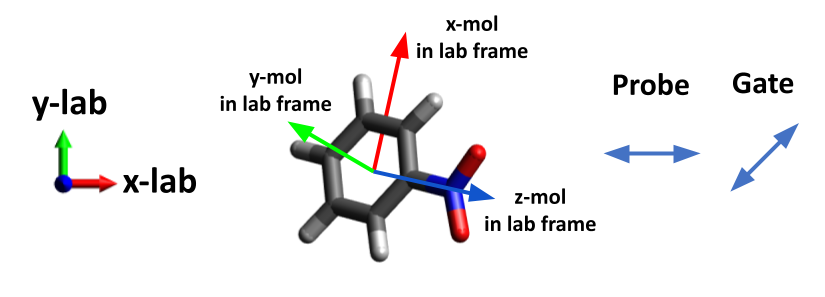}}
    \caption{A representative lab frame orientation of nitrobenzene recovered from a 3-level orientationally averaged third order projected density matrix calculation of nitrobenzene. In this calculation, the included states were $S_0$, $S_1$, and $S_2$ and the pulse duration was 15~fs with a central wavelength of 780~nm. We note that here that the lab frame axes are defined as follows: $\hat{z}_{\text{lab}}$ is the pulse propagation axis, $\hat{x}_{\text{lab}}$ is aligned along the probe polarization axis, and $\hat{y}_{\text{lab}}$ is the analyzer polarizer axis. The molecular frame axis $\hat{z}_{mol}$ is aligned along the CN bond axis, $\hat{y}_{mol}$ is aligned normal to the plane of the ring, and $\hat{x}_{mol}$ is orientated in the plane of the benzene ring.}
    \label{sec_sim:fig:recovered_mol_frame_cartoon_of_nb}
\end{figure}

\section{The Projected Density Density Matrix - Introduction, Justification, and Application}

At its core, the projected density matrix encodes information about both the quantum state of a model system and the orientational information of that system. The latter is included by combining the laboratory frame polarization measurement axis and the molecular frame density matrix through the inclusion of a unit vector $\mathbf{s}$ that represents the polarization analyzer measurement axis. The $\mathbf{s}$ vector is a molecular-frame representation of the laboratory frame of reference (Fig.~\ref{sec_sim:fig:recovered_mol_frame_cartoon_of_nb}), which is defined by the analyzer polarization axis, onto which polarization signals are projected. We define the projected density matrix as the tensor product of  $\mathbf{s}$ with the density matrix $\vert \rho \rangle \! \rangle$:

\begin{equation}
 \vert \vec{\rho}_{proj} \rangle \! \rangle  = \mathbf{s} \otimes \vert \rho \rangle \! \rangle
\end{equation}

The projected density matrix implicitly accounts for the molecular polarization projected onto the polarization analyzer, corresponding to a polarization-sensitive measurement. We apply the projected density matrix to find single molecular orientations that are good approximations of the orientationally averaged signal. The relationship between the polarization signals computed from the electronic polarization, the projected density matrix, and an approximate single molecule orientation is shown as follows

\begin{equation}
    \begin{aligned}
        \textrm{Signal}[ P ] & = 
        \textrm{Signal}[ \langle \! \langle \vec{\mu}_{mol} \vert \vec{\rho}_{proj} \rangle\rangle_{\text{ave}}] \\
       & \approx
        \textrm{Signal}[ 
        \langle \! \langle \vec{\mu}_{mol}    \vert \vec{\rho}(R_{\text{fit}}) \rangle  \! \rangle 
        ]
    \end{aligned}
    \label{eq:sec_PDM_math:Signal_relationships_P_rho_PDM}
\end{equation}
where $P$ is the time- and delay-dependent polarization of the system along the measurement axis, $\vec{\mu}_{mol}$ is the molecular frame dipole operator of the system, $(\cdot)_{ave}$ denotes orientational averaging, $\textrm{Signal}[\cdot]$ is a functional that computes the signal from $P$, and $R_{\text{fit}}$ is the rotation operator that defines the single molecule orientation that best approximates the orientationally averaged signal. This further provides information about the orientation of molecules producing simulated signals that are highly representative of the signals produced in the ensemble-average of randomly-oriented molecules, which can be retrieved and used to interpret experimental measurements that were performed in the laboratory frame. 

To find the representative orientation, we perform a fit that uses information from the orientationally averaged projected density matrix to constrain sampled orientations and find the best fit orientation using comparisons against single orientation calculations. This fit allows us to recover the single lab frame molecular orientation that best approximates the signal. To find the corresponding rotation operator ($R$), we first decompose the orientationally averaged projected density matrix into an approximate molecular frame polarization analyzer axis $\mathbf{s}_{\text{approx}}$ and a recovered density matrix $\vert \rho \rangle \! \rangle_{\text{approx}}$ in the following manner.

\begin{equation}
 \vert \vec{\rho}_{proj} \rangle \! \rangle_{\text{ave}}  \approx \mathbf{s}(R)_{\text{approx}} \otimes \vert \rho(R) \rangle \!\rangle_{\text{approx}}
\end{equation}

This decomposition suggests two different approaches to find $R$. 
In the first and simplest approach, we can use $\mathbf{s}_{\text{approx}}$ to constrain the orientation of the molecule and perform a 1-dimensional search to find the representative signal (Section~\ref{sec_PDM_math:sec:compute_R}). In the second approach, in principle we can use the recovered density matrix $\vert \rho \rangle \! \rangle_{\text{approx}}$ to perform a fit against the orientationally dependent density matrix $\vert \rho(R) \rangle \! \rangle$. Computationally, this can be more demanding than the first approach, as it involves recomputing the density matrix for a variety of orientations. The benefit of the second approach is it allows the direct determination of the rotation operator that best approximates the average signal without performing a search. In this work we primarily focus on the first method. 

In the following section (Section~\ref{sec_PDM_math:sec:justification}), we will derive the projected density matrix for a given lab frame analyzer polarization axis $\hat{\mathbf{e}}^{\text{pol}}$ and it's molecular frame counterpart $\mathbf{s}$ that defines a measurement axis for a specified polarization response. To accomplish this, we will start by considering the transformation of this polarization signal from the lab frame to the molecular frame. This transformation is at the heart of the projected density matrix and will be used to justify the projected density matrix as a means to encode the quantum mechanics of the system and to describe the resulting polarization signal. After deriving the projected density matrix in the general case, we will then explore how this object behaves in the context of perturbation theory (Section~\ref{sec_PDM_math:sec:perturbation}). In particular, we will show that, unlike the molecular-frame density matrix, the projected density matrix transforms like the signal it represents under orientational averaging. This allows us to retain density matrix information after the orientational averaging process. Using this result, we will then present an approximation method, where we determine a single molecular orientation that produces signals almost identical to the orientation-averaged signals (Sections~\ref{sec_PDM_math:sec:s}, \ref{sec_PDM_math:sec:perturbation}). We discuss applications of this framework to compute approximate orientationally averaged signals over longer time durations, by exploiting the computational cost savings gained from using single molecular orientation calculations (Section~\ref{sec:simulation}).


\subsection{Derivation and Justification of the Projected Density Matrix from Molecular Frame and Lab Frame Polarization Signals}\label{sec_PDM_math:sec:justification}

The relationship between a signal that is proportional to a dipole operator in the laboratory frame $\vec{\mu}_{lab}$ and the dipole operator in the molecular frame $\vec{\mu}_{mol}$ is essentially a rotation \cite{kwak_rigorous_2015, andrews_using_2004} from the molecular frame (defined by the $\hat{\mathbf{e}}^{mol}_{i}$ unit vectors) to the laboratory frame (defined by the $\hat{\mathbf{e}}^{lab}_{i}$ unit vectors). This rotation can be described using a rotation operator ($R$) as follows:

\begin{equation}
\label{eq:sec_PDM_math:mu_lab_to_mu_mol}
    \langle \! \langle \vec{\mu}_{lab} \vert \rho \rangle \rangle = R^{-1} \langle \! \langle \vec{\mu}_{mol} \vert \rho \rangle \rangle
\end{equation}

When performing a polarization sensitive measurement, the polarizer will define a lab-frame measurement axis $\hat{\mathbf{e}}^{pol}$ that will reject the dipole signal orthogonal to this axis. This effect can be added to Eq.~\ref{eq:sec_PDM_math:mu_lab_to_mu_mol} by simply taking the scalar product of $\hat{\mathbf{e}}^{pol}$ on both sides of the equation giving the following:

\begin{equation}
\label{eq:sec_PDM_math:mu_lab_to_mu_mol_pol_spec}
   \hat{\mathbf{e}}^{pol} \cdot \langle \! \langle \vec{\mu}_{lab} \vert \rho \rangle \rangle = \hat{\mathbf{e}}^{pol} \cdot R^{-1} \langle \! \langle \vec{\mu}_{mol} \vert \rho \rangle \rangle
\end{equation}

Expanding the right hand side (RHS) of Eq.~\ref{eq:sec_PDM_math:mu_lab_to_mu_mol_pol_spec} and collecting the terms under the trace yields

\begin{equation}
\label{eq:sec_PDM_math:mu_lab_to_mu_mol_pol_spec_raw_PDM}
    \hat{\mathbf{e}}^{pol} \cdot \langle \! \langle \vec{\mu}_{lab} \vert \rho \rangle \rangle = 
    Tr\left[ 
    \left( \mu^{mol}_x, \mu^{mol}_y, \mu^{mol}_z \right) \cdot
    \left( s_{x} \rho , s_{y} \rho , s_{z} \rho  \right)
     \right], 
\end{equation}

where the components $s_i$ represent the relative contributions of each component of the molecular frame electronic polarization to the lab frame polarization. The $s_i$ components represent a rotation of the lab-frame defined measurement axis $\hat{\mathbf{e}}^{pol}$ into the molecular frame, which is written explicitly as follows:

\begin{equation}
\label{eq:sec_PDM_math:projection_s_vector_def}
    s_{i} = \hat{\mathbf{e}}^{pol} \cdot R^{-1} \hat{\mathbf{e}}^{mol}_{i} \; \; \; ; \; \; \; i = x, y, z
\end{equation}

Eq.~\ref{eq:sec_PDM_math:mu_lab_to_mu_mol_pol_spec_raw_PDM}, combined with the definition of the components of $\mathbf{s}$ in equation \ref{eq:sec_PDM_math:projection_s_vector_def}, are then used to recover the definition for the projected density matrix:

\begin{equation}
\label{eq:sec_PDM_math:PDM_def}
    \vert \vec{\rho}_{proj} \rangle \! \rangle 
    = \begin{pmatrix}
        s_{x} \vert \rho \rangle \! \rangle \\
        s_{y} \vert \rho \rangle \! \rangle \\
        s_{z} \vert \rho \rangle \! \rangle 
    \end{pmatrix}
    = \mathbf{s} \otimes \vert \rho \rangle \! \rangle
\end{equation}

With this definition, we express measurable polarization signals in terms of the projected density matrix by rewriting the RHS of Eq.~\ref{eq:sec_PDM_math:mu_lab_to_mu_mol_pol_spec_raw_PDM}. This gives us the following expression for a polarization sensitive lab frame signal in terms of the projected density matrix:

\begin{equation}
    \label{eq:sec_PDM_math:signal_from_PDM_summary}
        \hat{\mathbf{e}}^{pol} \cdot \vec{P}_{\text{sig}} =
        \hat{\mathbf{e}}^{pol} \cdot \langle \! \langle  \vec{\mu}_{lab} \vert \rho \rangle \rangle =
        \langle \! \langle \vec{\mu}_{mol} \vert \vec{\rho}_{proj} \rangle \rangle 
\end{equation}

\subsection{Perturbation Theory and Orientational Averaging using the Projected Density Matrix}\label{sec_PDM_math:sec:perturbation}

Including orientational information in the projected density matrix has several useful consequences when considering orientationally averaged perturbative signals. Before we discuss these implications, we will first provide a brief review of time-dependent perturbation theory. The goal of this discussion is two-fold: first, we need to generate a set of expressions that will be used in our implementation of perturbative simulations later, and second, we aim to further justify the projected density matrix by exploring how, under rotational averaging, the perturbed projected density matrix will transform like the signal it represents. These properties result in an important consequence: upon rotational averaging, the 3\textsuperscript{rd}-order projected density matrix retains the quantum information about the model system.

In a typical molecular frame perturbative treatment of the n\textsuperscript{th}-order polarization response \cite{mukamel_principles_1995, boyd_non-linear_2008}, the n\textsuperscript{th}-order correction to the polarization in the time domain is expressed in terms of a time ordered integral expression, where the inner tensor product of an n\textsuperscript{th}-order response function $S^{(n)}_{mol}(t, \tau_n , ..., \tau_1)$ is taken with the appropriate number of interacting fields $\Vec{E}^{mol}_m$. Expressions for these response functions can furthermore be derived from the Liouville equation by considering a Dyson series expansion of a dipole perturbation in the interaction picture \cite{sakurai_modern_2011}. Expressions for the n\textsuperscript{th}-order molecular frame polarization response in terms of the classical response function and a quantum mechanical treatment are shown in Eqs.~\ref{eq:sec_PDM_math:mol_frame_pol_response_fn_orderN} and \ref{eq:sec_PDM_math:mol_frame_pol_qm_pert_theory_orderN}, respectively: 

\begin{widetext}
\begin{subequations}
\label{eq:sec_PDM_math:mol_frame_orderN_pol_eqs}
    \begin{equation}
    \label{eq:sec_PDM_math:mol_frame_pol_response_fn_orderN}
        \vec{P}^{(n)}_{mol} =
        \int^{t}_{t_0} d\tau_1 ... \int^{\tau_{n-1}}_{t_0} d\tau_n
        S^{(n)}_{mol}(t, \tau_n , ..., \tau_1)
        \prod^{n}_{m = 1} \vec{E}^{mol}_{m}(t, \tau_m)
    \end{equation}

    \begin{equation}
    \label{eq:sec_PDM_math:mol_frame_pol_qm_pert_theory_orderN}
        \vec{P}^{(n)}_{mol} =
        \left( \frac{-i}{\hbar} \right)^{n}
        \int^{t}_{t_0} d\tau_1 ... \int^{\tau_{n-1}}_{t_0} d\tau_n
        \left[
        \langle \! \langle \vec{\mu}_{mol} \vert
        \left(
        \mathcal{U}_0\left(\Delta T_f\right) 
        \prod^{n}_{m = 1} 
        \vec{\mu}_{mol} \mathcal{U}_0\left(\Delta \tau_{m}\right)
        \right)
        \vert \rho_0 \rangle\!\rangle
        \right]
        \prod^{n}_{m = 1} \vec{E}^{mol}_{m}
    \end{equation}
\end{subequations}
\end{widetext}

Here, we identify the n\textsuperscript{th}-order response function $S^{(n)}_{mol}$ as the bracketed term in Eq.~\ref{eq:sec_PDM_math:mol_frame_pol_qm_pert_theory_orderN}. $\mathcal{U}_0$ is the unperturbed time evolution operator in Liouville space, $\vec{\mu}_{mol}$ is the molecular frame dipole operator, $\vert \rho_0 \rangle\!\rangle$ is the unperturbed density matrix, and the amount of time between the interactions in perturbation theory is denoted by $\Delta \tau_{m}$ and $\Delta T_{f}$, which have the following definitions:

    \begin{equation*}
        \Delta T_f = t - \tau_1
    \end{equation*}
    
    \begin{equation*}
        \Delta \tau_m = 
        \begin{cases}
        \tau_m - \tau_{m + 1} ,& \text{if } m < n \\
        \tau_n - t_0,              & m = n 
        \end{cases}
    \end{equation*}


To translate the molecular frame polarization expressions of Eqs.~\ref{eq:sec_PDM_math:mol_frame_pol_response_fn_orderN} and \ref{eq:sec_PDM_math:mol_frame_pol_qm_pert_theory_orderN} into laboratory frame expressions \cite{kwak_rigorous_2015, andrews_using_2004} most relevant to practical simulations and experiments, we simply include the rotation operator $R$ that translates the lab frame measured fields $\Vec{E}^{lab}$ and polarization $\vec{P}^{(n)}_{lab}$ into their molecular frame counterparts. After applying this operation to both sides of Eq.~\ref{eq:sec_PDM_math:mol_frame_pol_qm_pert_theory_orderN} and solving for the lab frame response function $S^{(n)}_{lab}$, we get the following:

\begin{widetext}
\begin{subequations}
\label{eq:sec_PDM_math:lab_frame_orderN_pol_eqs}
    \begin{equation}
    \label{eq:sec_PDM_math:lab_frame_pol_response_fn_orderN}
        S^{(n)}_{lab} \equiv
        R^{-1}(\theta, \phi, \psi) S^{(n)}_{mol}(t, \tau_n , ..., \tau_1)
        \prod^{n}_{m = 1} R(\theta, \phi, \psi)
    \end{equation}

    \begin{equation}
    \label{eq:sec_PDM_math:lab_frame_resp_fn_qm_pert_theory_orderN}
        S^{(n)}_{lab} =
        R^{-1}(\theta, \phi, \psi) \langle \! \langle \vec{\mu}_{mol} \vert
        \left(
        \mathcal{U}_0\left(\Delta T_f\right) 
        \prod^{n}_{m = 1} 
        \vec{\mu}_{mol} R(\theta, \phi, \psi) \mathcal{U}_0\left(\Delta \tau_{m}\right)
        \right)
        \vert \rho_0 \rangle\!\rangle
    \end{equation}
\end{subequations}
\end{widetext}
These relations are then used to compute the lab frame n\textsuperscript{th}-order polarization response via the following expression:\cite{boyd_non-linear_2008, kwak_rigorous_2015, andrews_using_2004} 

\begin{equation}
    \label{eq:sec_PDM_math:lab_frame_pol_response_fn_orderN_long}
    \begin{aligned}
        \vec{P}^{(n)}_{lab} 
        \propto 
        \int^{t}_{t_0} d\tau_1 ... \int^{\tau_{n-1}}_{t_0} d\tau_n 
        S^{(n)}_{lab}
        \prod^{n}_{m = 1} \vec{E}^{lab}_{m}(t, \tau_m)
    \end{aligned}
\end{equation}

In order to perform a molecular frame to lab frame transformation of the perturbed density matrix, we follow a similar procedure. First, we consider the following equivalent expression for the n\textsuperscript{th}-order molecular frame polarization in terms of the n\textsuperscript{th}-order correction to the density matrix $\rho^{(n)}$\cite{mukamel_principles_1995}:
\begin{equation}
\label{eq:sec_PDM_math:mol_frame_polarization_orderN}
\vec{P}^{(n)}_{mol} =
        \langle \! \langle \vec{\mu}_{mol}
        \vert
        \rho^{(n)} \rangle\!\rangle
\end{equation}

Again, from perturbation theory \cite{mukamel_principles_1995}, the n\textsuperscript{th}-order correction to the density matrix, using optical fields defined in the laboratory frame, is expressed as follows:

\begin{widetext}
\begin{equation}
\label{eq:sec_PDM_math:rotated_density_matrix_orderN_long}
    \vert \rho^{(n)} (\theta, \phi, \psi)  \rangle\!\rangle 
    \propto
    \int^{t}_{t_0} d\tau_1 ... \int^{\tau_{n-1}}_{t_0} d\tau_n 
    \left(
    \mathcal{U}_0\left(\Delta T_f \right)
    \prod^{n}_{m = 1} \left( 
    \vec{\mu} R(\theta, \phi, \psi) \mathcal{U}_0\left(\Delta \tau_{m}\right) \right)
    \vert \rho_0 \rangle\!\rangle
    \right)
    \prod^{n}_{m = 1} \vec{E}^{lab}_{m},
\end{equation}
\end{widetext}

where the term in parentheses of Eq.~\ref{eq:sec_PDM_math:rotated_density_matrix_orderN_long} is roughly analogous to $S^{(n)}_{lab}$ (Eq.~\ref{eq:sec_PDM_math:lab_frame_resp_fn_qm_pert_theory_orderN}), except for the perturbation correction to the density matrix, as apposed to the perturbation correction to the polarization signal. 

In order to derive an expression for the perturbed projected density matrix, all that is necessary is to combine the results of Eq.~\ref{eq:sec_PDM_math:rotated_density_matrix_orderN_long} with the definition of the projected density matrix in Eq.~\ref{eq:sec_PDM_math:PDM_def}. This lets us express the n\textsuperscript{th}-order correction to the projected density matrix as follows:

\begin{widetext}
\begin{equation}
\label{eq:sec_PDM_math:nth_order_projected_density_matrix}
    \begin{aligned}
        \vert \vec{\rho}^{(n)}_{proj} \rangle \! \rangle = &
        \mathbf{s}(\theta, \phi, \psi) \otimes \vert \rho^{(n)}(\theta, \phi, \psi) \rangle \! \rangle
        = & 
         \begin{pmatrix}
            \hat{\mathbf{e}}^{pol} \cdot R^{-1}(\theta, \phi, \psi) \hat{\mathbf{e}}^{mol}_x \vert \rho^{(n)} (\theta, \phi, \psi) \rangle \! \rangle 
            \\
            \hat{\mathbf{e}}^{pol} \cdot R^{-1}(\theta, \phi, \psi) \hat{\mathbf{e}}^{mol}_y \vert \rho^{(n)} (\theta, \phi, \psi) \rangle \! \rangle 
            \\
            \hat{\mathbf{e}}^{pol} \cdot R^{-1}(\theta, \phi, \psi)
            \hat{\mathbf{e}}^{mol}_z \vert \rho^{(n)} (\theta, \phi, \psi) \rangle \! \rangle 
        \end{pmatrix}
    \end{aligned}
\end{equation}
\end{widetext}

There are a few observations to make at this stage. First, when comparing the expressions for the lab frame perturbed response function in Eq.~\ref{eq:sec_PDM_math:lab_frame_resp_fn_qm_pert_theory_orderN} to that of the equivalent response function term for the perturbed density matrix with lab frame defined fields in Eq.~\ref{eq:sec_PDM_math:rotated_density_matrix_orderN_long}, we find that these expressions are nearly identical, except for the lack of both the final contracting dipole operator and the corresponding inverse rotation operator in the perturbed density matrix in Eq.~\ref{eq:sec_PDM_math:rotated_density_matrix_orderN_long}. As such Eqs.~\ref{eq:sec_PDM_math:rotated_density_matrix_orderN_long} and \ref{eq:sec_PDM_math:lab_frame_resp_fn_qm_pert_theory_orderN} effectively represent two different expressions for the n\textsuperscript{th}-order response within the same perturbative treatment of the relevant system. 
After rotational averaging, the two orientationally averaged quantities of Eqs.~\ref{eq:sec_PDM_math:rotated_density_matrix_orderN_long} and \ref{eq:sec_PDM_math:lab_frame_resp_fn_qm_pert_theory_orderN} have different intrinsic permutation symmetries because of the different number of rotation operators. Specifically, the orientation-averaged density matrix (Eq.~\ref{eq:sec_PDM_math:rotated_density_matrix_orderN_long}) is almost always zero, even when the orientation-averaged signal (Eq.~\ref{eq:sec_PDM_math:lab_frame_resp_fn_qm_pert_theory_orderN}) is non-zero. After the inverse rotation operator is reintroduced in the n\textsuperscript{th}-order projected density matrix (Eq.~\ref{eq:sec_PDM_math:nth_order_projected_density_matrix}), one may expect that under orientational averaging the resulting projected density matrix should have the same intrinsic permutation symmetry as the signal it represents. This is most clearly illustrated when considering the intrinsic response function and the perturbed density matrix in the case of first order perturbation theory:

\begin{subequations}
    \begin{equation}
    \label{eq:sec_PDM_math:lab_frame_resp_fn_qm_pert_theory_order1}
        S^{(1)}_{lab}(R) =
        R^{-1} \langle \! \langle \vec{\mu}_{mol} \vert
        \mathcal{U}_0\left(\Delta T_f\right) 
        \vec{\mu}_{mol} R \mathcal{U}_0\left(\Delta \tau_{1}\right)
        \vert \rho_0 \rangle\!\rangle
    \end{equation}
    
    \begin{equation}
    \label{eq:sec_PDM_math:rotated_density_matrix_order1_pert_int_term}
        \vert \rho^{(1)} (R) \rangle\!\rangle_{i_1 ... i_n}
        = 
        \mathcal{U}_0\left(\Delta T_f \right)
        \vec{\mu} R \mathcal{U}_0\left(\Delta \tau_{1}\right)
        \vert \rho_0 \rangle\!\rangle       
    \end{equation}
\end{subequations}

In this case, the first order orientationally averaged response function $\langle S^{(1)}_{lab} \rangle$ will have nonzero terms along its diagonal elements \cite{boyd_non-linear_2008, kwak_rigorous_2015, andrews_using_2004}. However, the corresponding orientationally averaged first order correction to the density matrix $\langle \vert \rho^{(1)} (R) \rangle\!\rangle_{i_1 ... i_n} \rangle$, as it only contains one rotation operator, must be exactly zero, because the rotational average of a single static vector (as defined by $\vec{\mu}$) must be zero.

For the n\textsuperscript{th}-order projected density matrix (Eq.~\ref{eq:sec_PDM_math:nth_order_projected_density_matrix}), the elements of the perturbed projected density matrix transform like the signal they represent, due to the inclusion of the inverse rotation operator. This lets the projected density matrix retain dynamical information about the quantum evolution of the system after orientational averaging. In the following sections, we will show how we can use this dynamical information, which is retained after this transformation, to interpret results from simulations and greatly speed up time-dependent calculations.

\subsection{Recovering Molecular Frame Information using the Projected Density Matrix}\label{sec_PDM_math:sec:s}

By definition, $\mathbf{s}$ is a unit vector with real components describing the lab frame analyzer polarization axis in the molecular frame ( Eq.~\ref{eq:sec_PDM_math:projection_s_vector_def}). As such, $\mathbf{s}$ must have real components and unit magnitude, as shown in Eq.~\ref{eq:sec_PDM_math:mag_of_s_real_sense}:

\begin{equation}
\label{eq:sec_PDM_math:mag_of_s_real_sense}
   \vert \vert \mathbf{s} \vert \vert_{\text{real}}
   = \sqrt{
   \left( \Tilde{s}_x \right)^2 +
   \left( \Tilde{s}_y \right)^2 +
   \left( \Tilde{s}_z \right)^2
   }
   = 1
\end{equation}

However, because the elements of the density matrix are complex, the tensor elements of the projected density matrix must also in general be complex. Given the rich information contained in the projected density matrix, this presents an opportunity to over-determine $\mathbf{s}$ by recovering a set of measurement axes $\mathbf{s}^{\text{rec}}_{ij}$ for each real and imaginary component of the density matrix element contained in the given projected density matrix. By averaging these recovered vectors, we can recover the measurement axis $\mathbf{s}$ as well as confidence intervals associated with this vector up to an overall sign ambiguity, which is due to the double-ended nature of the polarization vector. In principle, this recovered measurement axis can then be used to determine the Euler angles that are associated with the orientation of the model molecule.

To start this analysis, first consider the following relationship between the elements of the projected density matrix and the corresponding orientational decomposition, as shown in Eq.~\ref{eq:sec_PDM_math:recovered_projected_density_matrix_elems}:

\begin{equation}
\label{eq:sec_PDM_math:recovered_projected_density_matrix_elems}
    \langle \! \langle i j \vert \vec{\rho}_{proj} \rangle \! \rangle
    =
    \mathbf{s} \langle \! \langle i j \vert \rho \rangle \! \rangle
    \approx
    \mathbf{s}^{\text{rec}} \langle \! \langle i j \vert \rho_{\text{approx}} \rangle \! \rangle
\end{equation}

In this equation, we see that the vector orientation comes purely from the analyzer polarization axis $\mathbf{s}$, whereas the complex magnitude of the vector stems from the density matrix element itself. This suggests that we can recover a vector approximating $\mathbf{s}$, using the following procedure. First, we separate the real and imaginary components of each density matrix element. As a result we get two separate real valued vectors that can be normalized into unit vectors. By normalizing these vectors, we isolate the real and imaginary parts, and recover the $\mathbf{s}^{\text{rec}}_{ij}$ vector up to a sign ambiguity for each density matrix element. This is summarized in Eqs.~(\ref{eq:sec_PDM_math:real_imag_recovered_projected_density_matrix_elems}a and b):
\begin{subequations}
\label{eq:sec_PDM_math:real_imag_recovered_projected_density_matrix_elems}
    \begin{equation}\label{eq:sec_PDM_math:real_recovered_projected_density_matrix_elems}
        \frac{
        \mathcal{R} \left[ \langle \! \langle i j \vert \vec{\rho}_{proj} \rangle \! \rangle \right]
        }{
        \left| \left|
        \mathcal{R} \left[ \langle \! \langle i j \vert \vec{\rho}_{proj} \rangle \! \rangle \right]
        \right| \right|_{\text{real}}
        }
        \equiv
        \pm
        \mathbf{s}^{\text{real}}_{ij}
    \end{equation}
    \begin{equation}\label{eq:sec_PDM_math:imag_recovered_projected_density_matrix_elems}
        \frac{
        \mathcal{I} \left[ \langle \! \langle i j \vert \vec{\rho}_{proj} \rangle \! \rangle \right]
        }{
        \left| \left|
        \mathcal{I} \left[ \langle \! \langle i j \vert \vec{\rho}_{proj} \rangle \! \rangle \right]
        \right| \right|_{\text{real}}
        }
        \equiv
        \pm
        \mathbf{s}^{\text{imag}}_{ij}
    \end{equation}
\end{subequations}

In principle, these two normalized vectors should recover the same vector quantity $\mathbf{s}_{ij}^{\text{ rec}}$. Additionally, for a single orientation, each density matrix element is expected to produce the same vectors $\mathbf{s}_{ij}^{\text{ rec}}$. As such, averaging the resulting recovered $\mathbf{s}_{ij}^{\text{ rec}}$ vectors over both the real and imaginary components of the sampled density matrix elements should give us a good representation of the molecular frame polarization analyzer axis $\mathbf{s}$. Additionally, we can compute the averaged recovered $\mathbf{s}^{\text{rec}}$ as well as confidence intervals from these recovered $\mathbf{s}_{ij}$ vectors. The final expression for $\mathbf{s}^{\text{rec}}$ from averaging the $\mathbf{s}_{ij}$ terms from both the real and imaginary components of the projected density matrix is:

\begin{equation}
\label{eq:sec_PDM_math:ave_s_from_sreals_simags}
    \frac{
    \langle 
    \mathbf{s}^{\text{imag}}_{ij}
    \rangle 
    +
    \langle 
    \mathbf{s}^{\text{real}}_{ij} 
    \rangle
    }{2} = \mathbf{s}^{\text{rec}}
\end{equation}

Lastly, factoring out $\mathbf{s}^{\text{rec}}$ from the given projected density matrix, as illustrated in Eq.~\ref{eq:sec_PDM_math:recovered_projected_density_matrix_elems}, allows us to compute an approximate, recovered density matrix up to an overall sign ambiguity. Now that we have a procedure for recovering the measurement axis $\mathbf{s}^{\text{rec}}$ and the density matrix $\vert \rho_{\text{approx}} \rangle$ from a given projected density matrix, we can next discuss an approximation method based on this procedure by which we compute the measurement axis and density matrix that best approximates the projected density matrix after orientational averaging.

\subsection{Approximating Orientationally Averaged Signals using a Single Molecular Frame Orientation}\label{sec_PDM_math:sec:compute_R}

Under orientational averaging, the projected density matrix can be written as: 

\begin{eqnarray}
\label{eq:sec_PDM_math:projected_density_matrix_rot_ave}
    \vert \vec{\rho}_{proj} \rangle \! \rangle_{ave} = &
    \int
    \mathbf{s}(R) \otimes \vert \rho(R) \rangle \! \rangle
    dR \nonumber
    \\
    = & \begin{pmatrix}
        \int
        \hat{\mathbf{e}}^{pol} \cdot R^{-1} \hat{\mathbf{e}}^{mol}_x \vert \rho (R) \rangle \! \rangle 
        dR
        \\
        \int
        \hat{\mathbf{e}}^{pol} \cdot R^{-1} \hat{\mathbf{e}}^{mol}_y \vert \rho (R) \rangle \! \rangle 
        dR
        \\
        \int
        \hat{\mathbf{e}}^{pol} \cdot R^{-1} \hat{\mathbf{e}}^{mol}_z \vert \rho (R) \rangle \! \rangle 
        dR
    \end{pmatrix}
\end{eqnarray}

Applying the results of perturbation theory to Eq.~\ref{eq:sec_PDM_math:projected_density_matrix_rot_ave} lets us write the n\textsuperscript{th}-order correction of the rotationally averaged projected density matrix as:

\begin{eqnarray}
\label{eq:sec_PDM_math:nth_order_projected_density_matrix_rot_ave}
    \begin{aligned}
        \vert \vec{\rho}^{(n)}_{proj} \rangle \! \rangle_{ave} = &
        \int
        \mathbf{s}(R) \otimes \vert \rho^{(n)}(R) \rangle \! \rangle
        dR 
        \\
        = & 
        \hat{\mathbf{e}}^{pol} \cdot \begin{pmatrix}
            \int
            R^{-1} \hat{\mathbf{e}}^{mol}_x \vert \rho^{(n)} (R) \rangle \! \rangle 
            dR
            \\
            \int
            R^{-1} \hat{\mathbf{e}}^{mol}_y \vert \rho^{(n)} (R) \rangle \! \rangle 
            dR
            \\
            \int
            R^{-1}
            \hat{\mathbf{e}}^{mol}_z \vert \rho^{(n)} (R) \rangle \! \rangle 
            dR
        \end{pmatrix}
    \end{aligned}
\end{eqnarray}

Here, we approximate the orientationally averaged projected density matrix as a single orientation projected density matrix, which is represented by the tensor product of a single molecular frame analyzer polarization axis (represented by $\mathbf{s}_{\text{approx}}$) and a single density matrix (represented by $\rho_{\text{approx}}$). Effectively, this assumes that there is a single dominant orientation of the molecule that contributes to a measured polarization selective signal. This can be justified as follows. Using time-dependent perturbation theory \cite{boyd_non-linear_2008, mukamel_principles_1995}, a signal is obtained from a finite sequence of specific interactions acting on transition dipole matrix elements. Since each transition will be maximized if the perturbing field is aligned along the transition dipole axis, this suggests that, for any given interaction sequence, there should be some molecular orientation that maximizes the projection of the perturbing fields with respect to the mediating transition dipoles and thus the resulting signal. While many such sequences of transitions can and do exist, the pathways that have major contributions to the net signal may go through a similar sequence of dipole transitions. If this is the case, the orientation that maximizes all of these contributions could be representative of the net orientationally averaged signal. We represent this approximation in the following equation:

\begin{equation}
\label{eq:sec_PDM_math:approx_projected_density_matrix_rot_ave}
    \vert \vec{\rho}_{proj} \rangle \! \rangle_{\text{ave}} =
    \int
    \mathbf{s}(R) \otimes \vert \rho(R) \rangle \! \rangle
    dR
    \approx
    \mathbf{s}_{\text{approx}} \otimes \vert \rho_{\text{approx}} \rangle \! \rangle
\end{equation}

Using the procedure outlined in the Section~\ref{sec_PDM_math:sec:s}, we recover both an approximate density matrix and the approximate orientation of the analyzer polarization axis $\mathbf{s}_{\text{approx}}$ in the molecular frame. Mathematically, this information by alone insufficient to fully determine the molecular orientation that best represents the orientationally averaged signal. This is because a rotation matrix in $\mathcal{R}^{3}$ is defined by three independent Euler angles, whereas a vector in $\mathcal{R}^{3}$ can be determined from only two Euler angles. One straightforward way to work around this limitation is to simply sample the signal from orientations that are constrained by the molecular frame orientation of the polarization axis. Effectively, this reduces a 3-dimensional search to a 1-dimensional search for the signal that best approximates the orientation-averaged calculation. To perform this search, we compute the molecular frame to lab frame rotation matrix by considering two separate axis-angle rotations prior to performing the perturbation calculation as follows

\begin{equation}
\label{eq:sec_PDM_math:rotations_around_s_vec}
    R_{\text{mol-to-lab}} = R_2(\hat{e}_{\text{orth}}, \theta_{\text{align}}) R_1(\mathbf{s}^{\text{approx}}, \theta_r)
\end{equation}

In Eq.~\ref{eq:sec_PDM_math:rotations_around_s_vec}  we rotate the molecule around the approximate molecular frame polarization analyzer axis $\mathbf{s}_{\text{approx}}$ by a variable amount $\theta_r$. Such rotations don't change the relative orientation of the polarization axis in the molecular frame, as this $\theta_r$ is the undetermined degree of freedom to be searched over. To align the molecular frame polarization analyzer axis to the lab frame polarization axis, we perform a second axis-angle rotation that aligns $\mathbf{s}_{\text{approx}}$ to the lab-frame polarization axis, which we have defined to be along the y-axis. This is done by finding the orthogonal axis $\hat{e}_{\text{orth}}$ between the y-axis and $\mathbf{s}_{\text{approx}}$ and rotating around the $\hat{e}_{\text{orth}}$ axis by the angle $\theta_{\text{align}}$ between these two unit vectors. Note that, because $\hat{e}_{\text{orth}}$ has a defined value, and because $\theta_{\text{align}}$ is determined from a calculation involving $\hat{e}_{\text{orth}}$ and $\mathbf{s}_{\text{approx}}$, the rotation matrix $R_2$ is effectively determined by $\mathbf{s}_{\text{approx}}$.

With this framework in place, in the next section we will derive these dominant molecular orientations by computing the exact orientationally averaged projected density matrix over a short real-time interval. Next, we will perform a fit to determine the orientation. We then perform molecular frame calculations over longer simulation time durations, using this recovered orientation as a low computational cost alternative to orientational averaging, to simulate the approximate signals produced by a sample of randomly oriented molecules.

\section{Perturbative Simulations of Orientationally Averaged Signals using the Projected Density Matrix Formalism}\label{sec:simulation}

We now aim to better understand how the electronic structure of nitrobenzene contributes to a simulated optical Kerr effect signal. We simulate a time resolved optical Kerr effect experiment, where the pump and probe pulses are linearly polarized with a relative angle of $45^{\circ}$ between them (see Fig.~\ref{sec_sim:fig:recovered_mol_frame_cartoon_of_nb}). The optical Kerr effect signal, orthogonal to the probe pulse, is produced by the transient nonlinear refractive index.\cite{mcmorrow_femtosecond_1988, palese_femtosecond_nodate} The signal is analyzed with a polarizer crossed at $90^{\circ}$ to the probe polarization to block the probe pulse. The time averaged polarization signal is measured as a function of the time-delay between the pump and probe pulses, resulting in a time delay dependent optical Kerr effect spectrum.

To model the electronic component of the polarization signal, we apply a damped N-level electronic model to describe the interaction of nitrobenzene with an incident electric field. This effectively isolates an electronic contribution to the optical Kerr effect signal. 

To parameterize the energy levels and dipole transitions of the model, we performed electronic structure calculations of nitrobenzene in a frozen nuclear geometry, using a Multiconfiguration Self Consistent Field (MCSCF) wavefunction method, as implemented in the Dalton quantum chemistry package \cite{aidas_dalton_2014} with the the dipole transitions of nitrobenzene being computed using a response theory method \cite{jonsson_response_1996}. The calculations were performed using a double zeta augmented Dunning correlation consistent basis with an active space consisting of 14 active electrons that are distributed among 11 orbitals. This active space was chosen based on previous electronic structure calculations that accurately described nondissociative relaxation mechanisms in nitrobenzene after excitation \cite{giussani_insights_2017}.

Prior gas phase measurements\cite{domenicano_molecular_1990} have found that the NO\textsubscript{2} functional group has a significant out-of-plane probability distribution, with an average dihedral angle of $13^{\circ}$ between the benzene ring and the NO\textsubscript{2} functional group. Therefore, in the present model, we optimize the ground state electronic structure in the average geometry. 

We compare two different electronic structure models, starting with a simple 3 level model, applying the ground electronic state $S_0$, and the lowest two singlet excited states $S_1$, and $S_2$. We also employ a 5 level model, which includes 5 singlet states $S_0$, $S_1$, $S_2$, $S_8$ and $S_{12}$. The latter two states are chosen for their computed strong transition dipole  moments. The decay and dephasing rates included in the time evolution operator $\mathcal{U}_0$ (Eq.~\ref{eq:sec_PDM_math:mol_frame_pol_qm_pert_theory_orderN}) are dependent on the electronic structure, and are all on the approximate order of magnitude ($\sim$100~fs) as measured in experiments \cite{lotshaw_femtosecond_1987, takezaki_relaxation_1998}.

\subsection{Simulation Results of Optical Kerr Effect Signals using the Projected Density Matrix Formalism}

\begin{figure}
    \centering
    \begin{tabular}{c }
             {\fontsize{0.4cm}{0.4cm}\selectfont a.
             
             }\\
    \subfloat{
    \includegraphics[width=0.7\columnwidth,trim={0.2cm 0.3cm 0.2cm 0.2cm},clip]{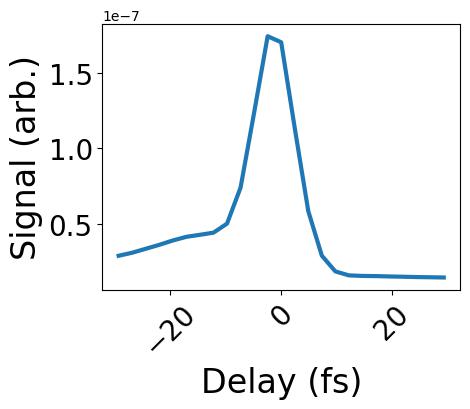}}
         \hfill
         \\
         {\fontsize{0.4cm}{0.4cm}\selectfont b.
             
             }
         \\
    \subfloat{     \includegraphics[width=0.7\columnwidth,trim={0.2cm 0.3cm 0.2cm 0.2cm},clip]{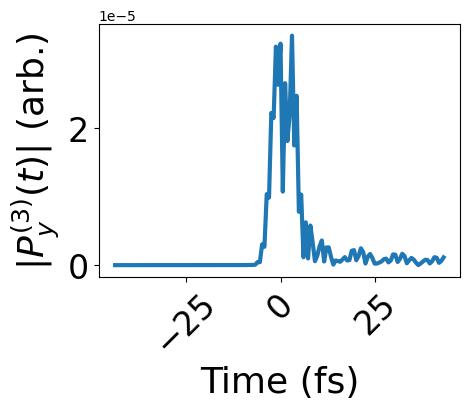}}
         
    \end{tabular}
    
    \caption{Simulated optical Kerr effect signal (a) as a function of time delay between two Gaussian transform-limited pulses with 7.5~fs duration and 780~nm central wavelength, and (b) third order polarization amplitude as a function of simulation time when the incident pulses are time-overlapped at 0~fs. The simulations employ the 3-level model described in Section~\ref{sec:simulation} and orientation averaging.}
    \label{sec_sim:fig:Simulated_averaged_signal_over_delay_and_polarization_over_time_3state}
\end{figure}

Using the nitrobenzene electronic structure model described in Section~\ref{sec:simulation}, a simulation of the polarization signal proceeds with the procedure summarized in Fig.~\ref{sec_sim:fig:workflow_for_single_run}. First we perform an orientationally averaged simulation (Step~1 in Fig.~\ref{sec_sim:fig:workflow_for_single_run}) over a limited simulation time window ($-44$ fs to $+44$ fs) which allows us to compute the orientationally averaged projected density matrix as a function of simulation time and time delay, shown in Fig.~\ref{sec_sim:fig:Simulated_averaged_signal_over_delay_and_polarization_over_time_3state}. 

We compute the orientationally averaged signal (Step~2a in Fig.~\ref{sec_sim:fig:workflow_for_single_run}) directly from the orientationally averaged projected density matrix of Eqs~\ref{eq:sec_PDM_math:Signal_relationships_P_rho_PDM} and \ref{eq:sec_PDM_math:signal_from_PDM_summary}. Through these calculations we compute a time dependent polarization for each time delay (Fig.~\ref{sec_sim:fig:Simulated_averaged_signal_over_delay_and_polarization_over_time_3state}b). From these polarization responses we compute a time delay dependent signal (Fig.~\ref{sec_sim:fig:Simulated_averaged_signal_over_delay_and_polarization_over_time_3state}a). This time delay dependent signal is what will be used as the basis for comparison in the later fitting steps. 

\begin{figure}[h!]
    \includegraphics[width=0.7\columnwidth]{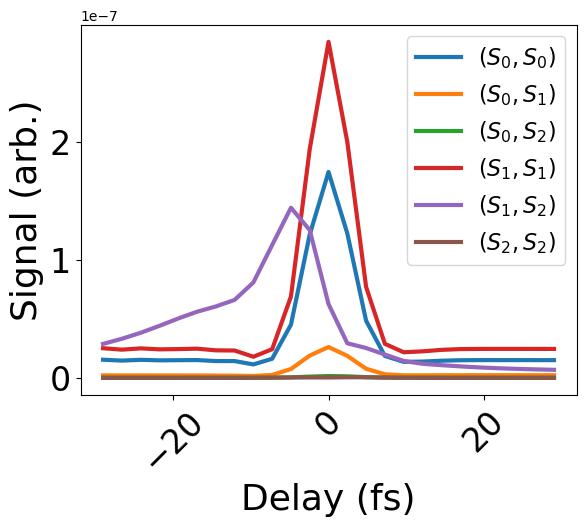}
    
    \caption{The orientationally averaged optical Kerr effect signal for each density matrix element as a function of time delay as computed from the orientationally averaged projected density matrix.}
    \label{sec_sim:fig:aved_signals_by_density_matrix_element_3state}
\end{figure}

In Fig.~\ref{sec_sim:fig:aved_signals_by_density_matrix_element_3state} we show the signals separated by density matrix element, offering insight into how the density matrix elements contribute to the computed signal after orientational averaging. From these calculations we can see that the three most important density matrix elements in the 3-state simulation as ordered by peak height are the ($S_1$, $S_1$) and ($S_0$, $S_0$) populations, followed by the ($S_1$, $S_2$) coherence with asymmetry in the signal primarily coming from the ($S_1$, $S_2$) coherence.

\begin{figure*}[]
    {\includegraphics[width=0.70\linewidth,trim={0.2cm 0.2cm 0.2cm 0.5cm},clip]{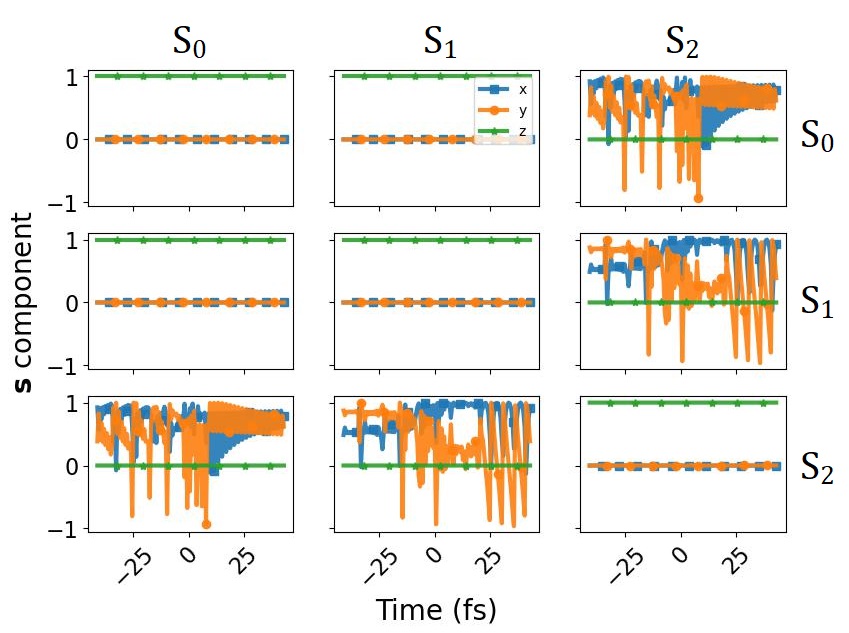}}
    
    \caption{Vector components of the molecular frame polarization analyzer axis $\mathbf{s}(t,T_{delay},\rho_{ij})$ as a function of simulation time. Each density matrix element is shown in a separate panel. The simulation employs two 7.5~fs IR pulses with 0 delay, and the 3-level model described in Section~\ref{sec:simulation}. Here we show the component values of $\mathbf{s}$  using only the real component of the projected density matrix. To find the averaged $\mathbf{s}$ we use calculated molecular frame polarization axes $\mathbf{s}$  from both the real and imaginary components of $\mathbf{s}$.}
    \label{sec_sim:fig:raw_s_vectors_overlapped pulses_3state}
\end{figure*}

From the orientationally averaged calculations, we then compute the molecular frame polarization analyzer axis $\mathbf{s}$ (Step~2b1 in Fig.~\ref{sec_sim:fig:workflow_for_single_run}) as a function of simulation time, time delay, and density matrix element for both the real and imaginary components of the projected density matrix. The results of a 3-level calculation are shown in Fig.~\ref{sec_sim:fig:raw_s_vectors_overlapped pulses_3state}, where we plot the component values of $\mathbf{s}$ as a function of simulation time for the case of overlapped pump and probe pulses, both of which are Gaussian transform limited 7.5 fs pulses with a center frequency of 780 nm. Note the recovered $\mathbf{s}$ components are shown using only the real component of the projected density matrix but later steps in the workflow of Fig.~\ref{sec_sim:fig:workflow_for_single_run} also use $\mathbf{s}$ as computed from the imaginary component of the projected density matrix.

Next, from the recovered molecular frame analyzer polarization $\mathbf{s}(t,T_{delay},\rho_{ij})$, an average $\mathbf{s}_{ave}$ is computed (Step~2b2 in Fig.~\ref{sec_sim:fig:workflow_for_single_run}) using the steps outlined in section~\ref{sec_PDM_math:sec:s}. Here we perform calculations using both our 3-state model with the computed $S_0$, $S_1$, and $S_2$ singlet states of nitrobenzene, and a 5-state model using states $S_0$, $S_1$, $S_2$, $S_8$, and $S_{12}$. The results from each of the two simulations are shown in Table~\ref{sec_sim:tab:aved_recovered_s_vectors}.

\begin{table}[h]
    \centering
    \begin{tabular}{c | c c c}
         & $s_x$ & $s_y$ & $s_z$ 
         \\ \hline
         3 state
         &  $0.3643 \pm 0.0034$ & $0.1980 \pm 0.0080$ & $0.527736 \pm 0.000030$ \\
         5 state 
         &  $0.3448 \pm 0.0026$ & $0.1791 \pm 0.0066$ & $0.509887 \pm 0.000033$ \\
    \end{tabular}
    \caption{Recovered components of the averaged molecular frame polarization analyzer axis $\mathbf{s}$  and standard errors from third order orientationally averaged projected density matrix calculations, using a pair of Gaussian transform-limited pulses with 7.5~fs duration and 780~nm central wavelength.}
    \label{sec_sim:tab:aved_recovered_s_vectors}
\end{table}

\begin{figure*}[]
    \centering
    \begin{tabular}{c c}
         a. \{$S_0$, $S_1$, $S_2$\} & b. \{$S_0$, $S_1$, $S_2$\} \\
         \resizebox{0.5\linewidth}{!}{\includegraphics{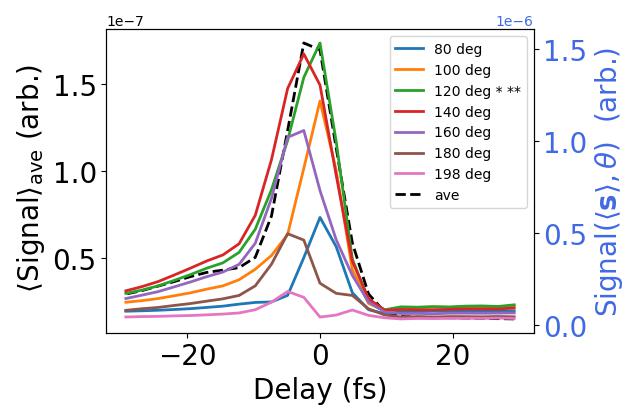}}
         &
         \resizebox{0.5\linewidth}{!}{\includegraphics{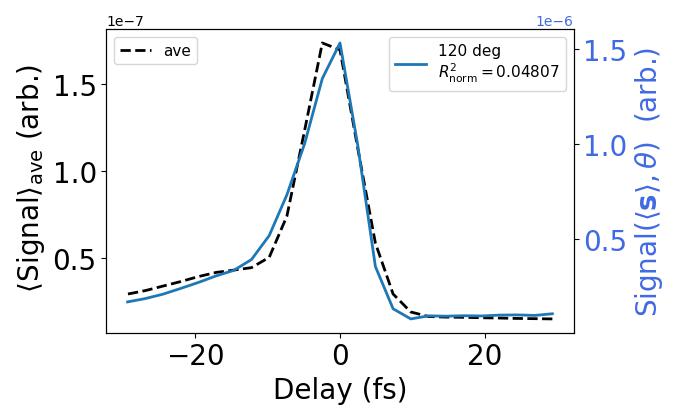}} \\
         c. \{$S_0$, $S_1$, $S_2$, $S_8$, $S_{12}$\} & d. \{$S_0$, $S_1$, $S_2$, $S_8$, $S_{12}$\} \\
         \resizebox{0.5\linewidth}{!}{\includegraphics{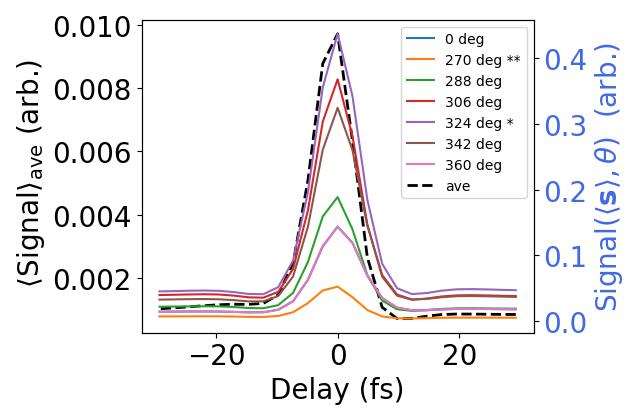}}
         &
         \resizebox{0.5\linewidth}{!}{\includegraphics{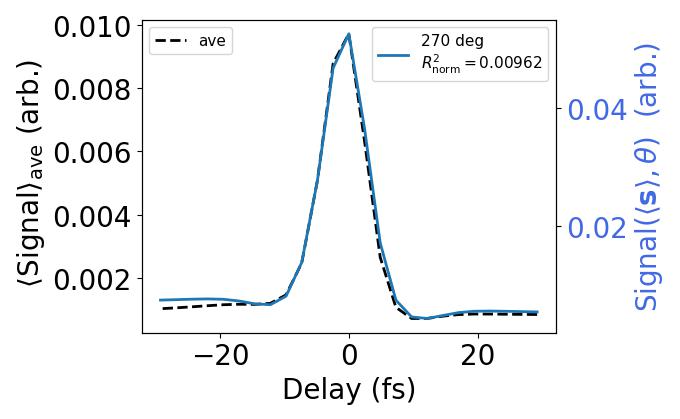}} 
    \end{tabular}
    
    \caption{Results from sampling the signal over a range of Euler angles $\theta_r$ about the retrieved  molecular frame polarization analyzer axis $\mathbf{s}$  (Table~\ref{sec_sim:tab:aved_recovered_s_vectors}). In panels a. and c. a single star (*) label denotes the curve with the sampled signal with the highest peak value, and two stars (**) label the sampled curve with the best fit as determined by the smallest normalized sum of square residuals value ($R^2_{\text{norm}}$) as defined in Eq.~\ref{eq:sec_PDM_math:R2norm_Fit_eval}. Here we sampled $\theta_r$ with a step size of roughly 18 degrees and show only signals with peak signals within an order of magnitude of the largest sampled signal.  Panels b. and d. show the orientation-averaged signal and the best $R^2_{norm}$ of the sampled orientations.}
    \label{sec_sim:fig:sample_signals_around_s_vectors}
\end{figure*}

We compute the time-averaged signal in Step~2b3 of Fig.~\ref{sec_sim:fig:workflow_for_single_run} from the average molecular frame analyzer polarization $\mathbf{s}_{ave}$. In Step~3, using the orientationally averaged signal from Step~1, and by using the averaged $\mathbf{s}_{ave}$ to constrain the search space (Sec.~\ref{sec_PDM_math:sec:compute_R}), we evaluate the resulting time delay dependent signals by using the following fit function:

\begin{equation}
\label{eq:sec_PDM_math:R2norm_Fit_eval}
    R^{2}_{norm} = \sum \left( \| \left<\text{Signal}\right>\| - \| \, \text{Signal} \! \left(\ \!\! \mathbf{s}_{\text{ave}}, \theta_r \right) \| \right)^2
\end{equation}

This evaluation function is chosen to compare the orientationally averaged signal with a single orientation calculation, and with the appropriate normalization. In the orientationally averaged calculation, many orientations produce weak contributions to the polarization response. By contrast, in the single orientation calculation, those weak contributions are excluded, resulting in single orientation delay-dependent signals with peak values that are significantly larger than the orientationally averaged signal. To account for the different signal scales in the evaluation function, first we normalize the single orientation signal, and the orientationally averaged signal, then take the $R^2$ from those normalized singles giving us the $R^2_{norm}$ which we use to evaluate the quality of the fits.

Sampling the remaining undetermined Euler angle $\theta_r$ as defined in Eq.~\ref{eq:sec_PDM_math:rotations_around_s_vec} from 0 to $2\pi$ yields the results of Fig.~\ref{sec_sim:fig:sample_signals_around_s_vectors} with the $\theta_r$ producing the largest amplitude and the best fit denoted by * and **, respectively. Using the angle $\theta_r$ giving the best fit, we recover a lab frame representative orientation of the nitrobenzene model as shown in Fig.~\ref{sec_sim:fig:recovered_mol_frame_cartoon_of_nb}.

We find that the simulation with the fixed orientation allows significant computational cost savings compared to the simulation employing explicit orientational averaging (Fig.~\ref{sec_sim:fig:aved_signals_by_density_matrix_element_3state}). Specifically, for the present 3-state nitrobenzene model, we find a computational cost saving of 1000 for the fixed orientation simulations compared to orientation averaging. For the 5-level nitrobezene model, we find a similar if somewhat smaller computational cost saving of 500. As expected, at the level of third order time-dependent perturbation theory where the computational cost scales with $t^3$. As such, we find the simulations employing the 3-level model and the fixed orientation have the same computational cost as the orientationally averaged calculation with a factor of 10 shorter time window.

\begin{figure}
    \centering
    \begin{tabular}{c }
             {\fontsize{0.4cm}{0.4cm}\selectfont a.
             
             }\\
    \subfloat{
    \includegraphics[width=0.8\columnwidth,trim={0.2cm 0.15cm 0.2cm 0.2cm},clip]{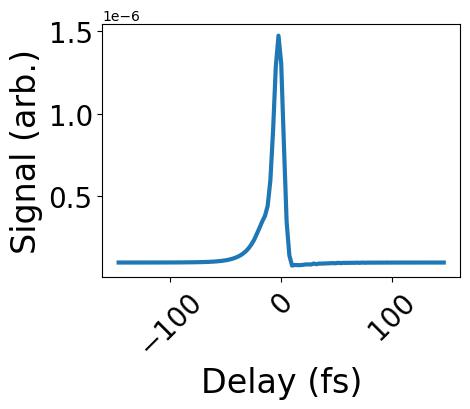}}
         \hfill
         \\
         {\fontsize{0.4cm}{0.4cm}\selectfont b.
             
             }
         \\
    \subfloat{     \includegraphics[width=0.8\columnwidth,trim={0.2cm 0.3cm 0.2cm 0.2cm},clip]{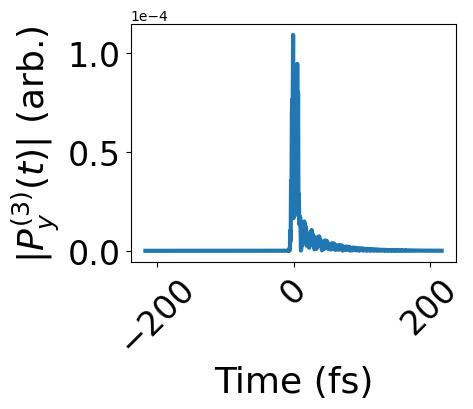}}
         
    \end{tabular}
    
    \caption{Simulated optical Kerr effect signal (a) as a function of time delay between two Gaussian transform-limited pulses with 7.5~fs duration and 780~nm central wavelength, and (b) third order polarization amplitude as a function of simulation time when the incident pulses are time-overlapped at 0~fs. The simulation employ the 3-level model described in Section~\ref{sec:simulation}. and the representative orientation retrieved by the procedure described in Section~\ref{sec_PDM_math:sec:compute_R}.}
    \label{sec_sim:fig:Simulated_oriented_signal_over_delay_and_polarization_over_time_3state}
\end{figure}

With this representative orientation approximately describing a sample of randomly oriented molecules, we arrive at Step~4 in the workflow of Fig.~\ref{sec_sim:fig:workflow_for_single_run} and simulate the over an extended simulation time window (-218 fs to 218 fs) and an extended time delay window (-146 fs to 146 fs). The results of this simulation are presented in Fig.~\ref{sec_sim:fig:Simulated_oriented_signal_over_delay_and_polarization_over_time_3state}. The simulation time-dependent polarization is found to decay completely in about 150~fs, which is well within the longer time range. The delay dependent signal decays sharply with the 7.5~fs pulse overlap just after 0 delay, but rises more slowly between delays of -100~fs and overlap. When fitting this negative delay rise to an exponential, we retrieve a time constant of 12.7 fs which is consistent with the set 13.1 fs dephasing time constant that was set for the ($S_1$, $S_2$) coherence.

When we compare this delay- and time-dependence with the fixed orientation results of Fig.~\ref{sec_sim:fig:Simulated_averaged_signal_over_delay_and_polarization_over_time_3state}, we note a good qualitative agreement between Fig.~\ref{sec_sim:fig:Simulated_oriented_signal_over_delay_and_polarization_over_time_3state} and the orientationally averaged signal over shorter simulation times in Fig.~\ref{sec_sim:fig:Simulated_averaged_signal_over_delay_and_polarization_over_time_3state}. In particular we note how in both the orientationally averaged and in the longer fixed orientation simulation the rise time of the signal in the simulation appears to be due to signal coming from the ($S_1$, $S_2$) coherence (see Fig.~\ref{sec_sim:fig:aved_signals_by_density_matrix_element_3state}) as seen in the decomposition of the signals into the contributions from individual density matrix elements. We also not that the rise time is consistent with the the specified dephasing rate for that coherence.

\section{Conclusions}

We have presented a method to recover molecular frame information for a sample of randomly-oriented molecules emitting nonlinear optical spectroscopic signals. The method involves the projection of a polarization analyzer axis onto the density matrix, connecting the laboratory frame to the molecular frame. This in turn allows the identification of a molecular orientation that generates approximately the same signal as is achieved through orientation averaging. We demonstrated the approach using an electronic model of nitrobenzene and optical Kerr effect simulations, while employing third order time-dependent perturbation theory. While we have developed and applied the method to time-dependent perturbation theory, we anticipate the method to be generally applicable to both perturbative and nonperturbative calculations of nonlinear spectroscopic signals. This computational cost-saving alternative to orientation averaging is expected to enable new time-domain nonlinear spectroscopy simulations of randomly-oriented molecules in the gas phase, liquids, and solutions. 

\section{Acknowledgments}
Work at LBNL was supported by the US Department of Energy (DOE), Office of Science (Sc), Division of Chemical Sciences Geosciences and Biosciences (CSGB) of the Office of Basic Energy Sciences (BES) under
Award No. DE-AC02-05CH11231. Data analysis was provided by a user project at the Molecular Foundry, supported by the DOE, Sc, BES under Contract No. DE-AC02-05CH11231. NS acknowledges support from the DOE, Sc, BES, CSGB under Award No. DE-SC0024234.

\section*{References}
\bibliography{PDM_references}

\end{document}